\documentclass[a4paper,11pt]{article}
\usepackage{pos}
\usepackage{amsmath}
\usepackage{graphbox}
\usepackage[version=4]{mhchem}

\title{Nuclear Matrix Elements for Neutrinoless Double-Beta Decay}
\ShortTitle{Neutrinoless Double-Beta Decay}

\author*[a]{Anthony V. Grebe}

\affiliation[a]{Fermi National Accelerator Laboratory,\\
Kirk and Pine Street, Batavia, IL, USA}

\emailAdd{agrebe@mit.edu}

\abstract{Neutrinoless double-beta decay ($0\nu\beta\beta$) is a rare hypothesised process that, if discovered, would establish that the neutrino is Majorana, that is, it is its own antiparticle.  Interpretation of experimental results relies on knowledge of nuclear matrix elements, whose large model uncertainty is the limiting factor in comparing measured (bounds on) half-lives to the neutrino mass.  Nuclear effective field theory and lattice QCD have the potential to compute these matrix elements with better control over uncertainties, enhancing the discovery potential of next-generation $0\nu\beta\beta$ experiments.  This work will survey various lattice QCD double-beta decay calculations and discuss their implications.}

\FullConference{The 41st International Symposium on Lattice Field Theory (LATTICE2024)\\
 28 July - 3 August 2024\\
Liverpool, UK\\}

\begin{document}
\maketitle

\section{Introduction}
The ghostly neutrino is the most poorly understood particle within the Standard Model.  In the original formulation of the Standard Model, neutrinos were assumed to be massless and left handed.  Experimental measurements, beginning with the Homestake experiment \cite{homestake} and its successors \cite{super-kamiokande, sno} have definitively shown that neutrinos oscillate, necessitating small but nonzero masses for at least two of the energy eigenstates.

The most straightforward extension to the minimal Standard Model to account for neutrino masses is to couple the neutrinos to the Higgs boson, analogously to the other massive fermions, as \cite{schwartz}
\begin{equation}
  \mathcal{L}_\text{Yukawa} = -Y_{ij}^e \bar L^i H e_R^j - Y_{ij}^\nu \bar L^i \tilde{H} \nu_R^i + \text{\emph{h.c.}}
  \label{neutrino-yukawa}
\end{equation}
with $L_i = \left( \begin{matrix} \nu_i \\ i \end{matrix} \right)$ the lepton doublet for lepton flavour $i$, $H = \frac{1}{\sqrt{2}} \left( \begin{matrix} 0 \\ v+h \end{matrix} \right)$ the Higgs doublet with vacuum expectation value $v = 251$ GeV, and $Y^e, Y^\nu$ matrices of Yukawa couplings.  Such terms require the introduction of right-handed neutrinos $\nu_R$ neutral under all Standard Model charges.  To produce sub-eV neutrino masses from electroweak-scale physics, however, requires Yukawa couplings $Y_{ij}^\nu \sim m_\nu / v \lesssim 10^{-12}$, at least six orders of magnitude smaller than that for the electron.

An alternative explanation is to postulate that the right-handed neutrino is Majorana, that is, $\nu_R = \bar \nu_R$, and to introduce a mass term of the form \cite{majorana}
\begin{equation}
  \mathcal{L}_\text{Majorana} \supset -iM_{ij} \left( \nu_R^i \right)^c \nu_R^j \, ,
  \label{majorana-mass}
\end{equation}
which violates none of the charge conservation laws due to the neutrality of $\nu_R$.  Combining this with the mass generated by the Yukawa couplings (in a simple 1-flavour model) gives a total mass
\begin{equation}
  \mathcal{L}_\text{mass} = -m \bar \nu_L \nu_R - M \bar \nu_R \nu_R + \text{\emph{h.c.}} = \left( 
    \begin{matrix}
      0 & m \\
      m & 2M
    \end{matrix}
  \right)
  \label{total-mass}
\end{equation}
with $m = yv/\sqrt{2}$.  Diagonalizing the mass matrix gives eigenvalues $\sqrt{m^2 + M^2} \pm M \sim \left\{ m^2/M, M \right\}$ when $m \ll M$.  With a Yukawa coupling $y$ of order 1 so that $m \sim 100$ GeV and a GUT- or Planck-scale Majorana mass $M \gtrsim 10^{16}$ GeV, one naturally obtains sub-eV scale masses for the eigenstate that overlaps primarily with the left-handed neutrino \cite{schwartz}.

The mass term in Eq.~(\ref{majorana-mass}) violates lepton number by two units, so experimental searches for its effects are limited to processes involving pairs of neutrinos.  Within the Standard Model, two antineutrinos can be produced in the lepton-number-conserving process of double-beta decay, involving the simultaneous beta decay of two neutrons into two protons,
\begin{equation}
  nn \rightarrow pp ee \bar \nu_e \bar \nu_e \, .
  \label{neutrinoful-double-beta-decay}
\end{equation}

As a second-order weak process, the reaction rate for neutrinoful double-beta decay ($2\nu\beta\beta$) is extremely slow, with half-lives exceeding the age of the universe by at least ten orders of magnitude.  As such, it is all but impossible to measure in nuclides that undergo the first-order single-beta decay reaction $n \rightarrow p e \bar \nu_e$, so experimental observation is limited to nuclides in which single-beta decay is forbidden.

\begin{figure}[h]
  \centering
  \includegraphics[width=0.49\textwidth]{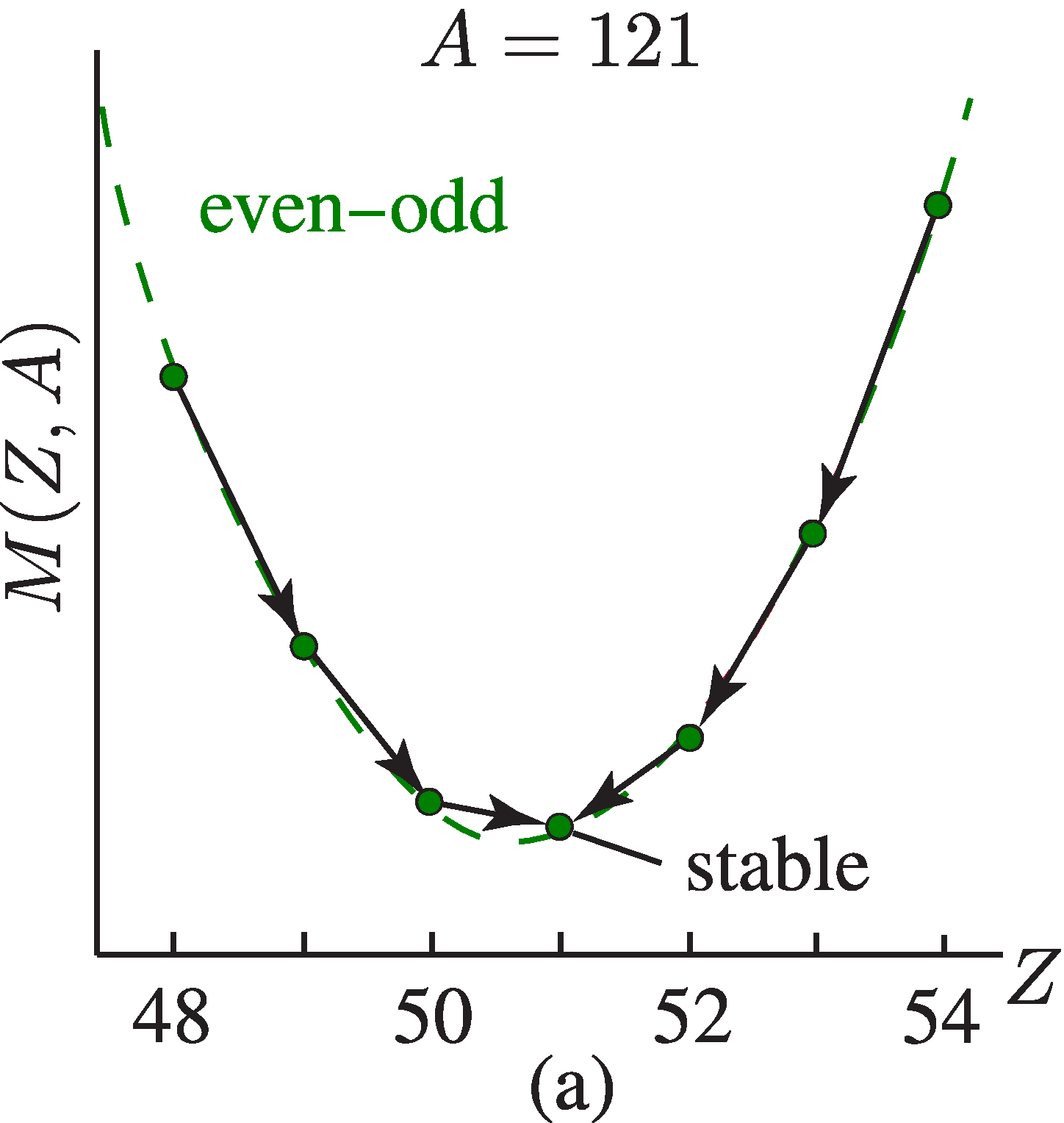}
  \includegraphics[width=0.49\textwidth]{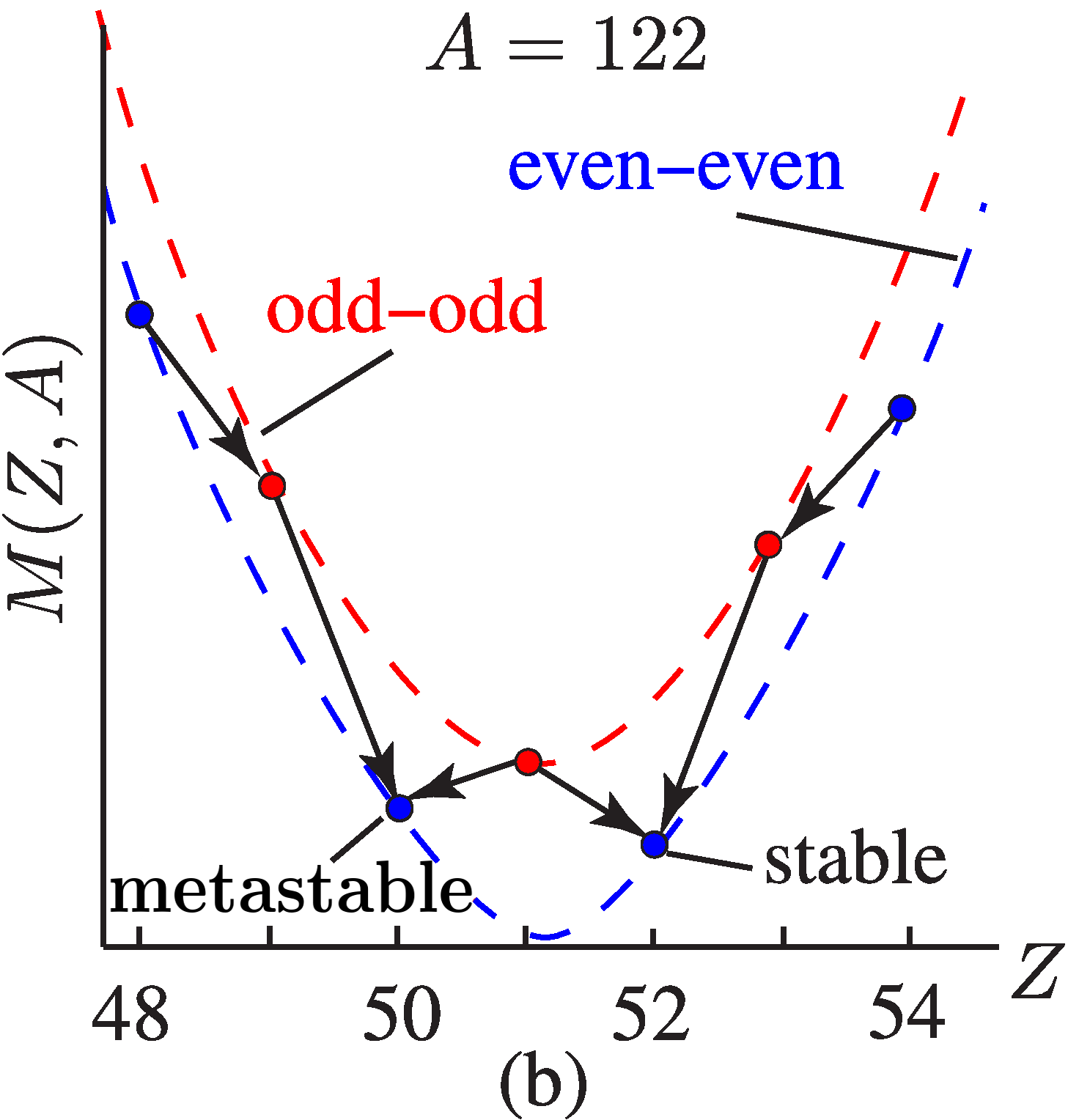}
\caption{(Left) For an odd-$A$ isobar, the semi-empirical mass formula in Eq.~(\ref{semi-empirical-mass}) predicts the mass is a quadratic function of $Z$ with a single minimum-mass nuclide (here \ce{^{121}_{51}Sb}) towards which all others $\beta$ decay.  (Right) for an even-$A$ isobar, the pairing term in Eq.~(\ref{pairing-term}) introduces oscillations in $Z$ between higher-energy odd-odd and lower-energy even-even parabolas, leading to non-convex and even non-monotonic behaviour.  As a result, there can be multiple local minima stable against single $\beta$ decay.  However, the higher-energy local minimum \ce{^{122}_{50}Sn} is unstable against the double-beta decay reaction in Eq.~(\ref{neutrinoful-double-beta-decay}), which would transition it to the lower-energy global minimum \ce{^{122}_{52}Te}, which is absolutely stable.  Figure adapted from Ref.~\cite{jaffe-taylor} after Ref.~\cite{jaffe-taylor-source}.}
  \label{isobar-masses}
\end{figure}

To a reasonable approximation, the mass of an nuclide with $Z$ protons and $N = A-Z$ neutrons is given by the semi-empirical mass formula \cite{jaffe-taylor}
\begin{equation}
  M(Z, A) = (m_p Z + m_n N) - \varepsilon_V A + \varepsilon_S A^{2/3} + \varepsilon_C \frac{Z^2}{A^{1/3}} + \varepsilon_\text{sym} \frac{(N-Z)^2}{A} + \eta (Z,N) \frac{\Delta}{A^{1/2}} \,
  \label{semi-empirical-mass}
\end{equation}
with positive constants $\varepsilon_V, \varepsilon_S, \varepsilon_C, \varepsilon_\text{sym}, \Delta$ fit to experimental data,
which at fixed atomic number $A$ is quadratic in $Z$ except for the pairing term
\begin{equation}
  \eta(Z, N) = 
   \left\{ 
    \begin{matrix}
      +1 & Z, N \text{ both odd} \\
      -1 & Z, N \text{ both even} \\
      0 & \text{otherwise}
    \end{matrix}
  \right. 
  \label{pairing-term}
\end{equation}
that favours nuclides with even numbers of protons and neutrons.  As a result, while odd-$A$ isobars (functions of constant $A$) are convex functions of $Z$ with a single minimum, masses of even-$A$ nuclides oscillate as a function of $Z$ between two parabolas split by a couple MeV, leading to non-monotonic behaviour in $Z$ and multiple local minima (see Fig.~\ref{isobar-masses}).
As the single-$\beta$ decay process is energetically forbidden, double-$\beta$ decay offers the only route from the metastable local minimum to the true global minimum.
The $2\nu\beta\beta$ reaction in Eq.~(\ref{neutrinoful-double-beta-decay}) was hypothesised by Maria Goppert-Meyer in 1935 \cite{mayer-double-beta}, discovered in 1950 from geological analysis \cite{double-beta-discovery}, and directly observed in laboratory settings in 1987 in \ce{^{82}_{34}Se} \cite{double-beta-direct-observation} and more recently in about a dozen other nuclides \cite{double-beta-status-prospects}.

If the neutrino is Majorana, the antineutrinos produced in Eq.~(\ref{neutrinoful-double-beta-decay}) can mutually annihilate, leaving just two electrons and two protons in the final state,
\begin{equation}
  nn \rightarrow pp ee \, .
  \label{neutrinoless-double-beta-decay}
\end{equation}
In this neutrinoless double-beta decay ($0\nu\beta\beta$) reaction, the entire decay energy is carried by the electrons (and, to a much lesser extent, by the recoiling nucleus) and can be observed in an experimental detector as a monoenergetic spike at the tail of the $2\nu\beta\beta$ spectrum.  Experimental searches for $0\nu\beta\beta$ have been carried out with a variety of nuclides \cite{gerda, cuore, kamland-zen}, placing increasingly stringent bounds on $0\nu\beta\beta$ half-lives $T_{1/2}^{0\nu}$, the longest being $3.8 \times 10^{26}$ years from KamLAND-Zen \cite{kamland-zen, kamland-zen-updated}.
Next-generation $0\nu\beta\beta$ experiments to refine this bound and to place stringent bounds on other nuclides have been identified as one of the highest priorities in the United States Department of Energy's Long-Range Plan for Nuclear Science \cite{long-range-plan}.

\begin{figure}[h]
  \centering
  \includegraphics{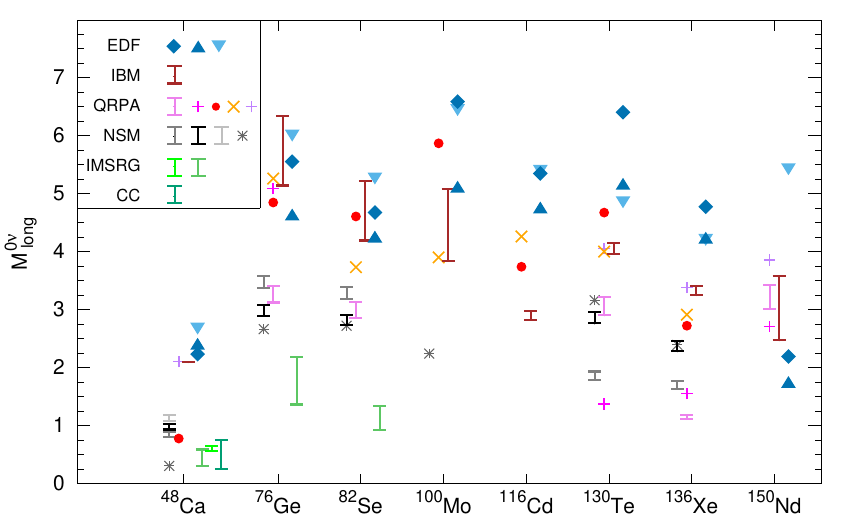}
  \caption{Various models used to predict the $0\nu\beta\beta$ matrix elements $M^{0\nu}$ for experimentally relevant nuclides.  Figure taken from Ref.~\cite{nuclear-matrix-elements-plot}.}
  \label{nme-models}
\end{figure}

In the minimal extension to the Standard Model given by Eq.~(\ref{majorana-mass}) above,
the experimental constraints can be related to the effective double-beta neutrino mass
\begin{equation}
  m_{\beta\beta} = \left| \sum_k U_{ek}^2 m_k \right| \, ,
  \label{double-beta-mass}
\end{equation}
where $m_k$ are the three light neutrino mass eigenstates and $U_{ek}$ is the Pontecorvo-Maki-Nakagawa-Sakata matrix \cite{pontecorvo, mns}, by
\begin{equation}
  \left( T_{1/2}^{0\nu} \right)^{-1} = |m_{\beta\beta}|^2 G^{0\nu} |M^{0\nu}|^2 \, ,
  \label{decay-relation}
\end{equation}
where $G^{0\nu}$ is a known kinematical factor depending on the decay energy and $M^{0\nu}$ is the nuclear matrix element encoding the structure of the parent and daughter nuclides.  Estimates for $M^{0\nu}$ are made using various nuclear models and, for a given nuclide, vary by a factor of several (see Fig.~\ref{nme-models}).  This model uncertainty introduces a corresponding uncertainty in the extracted value of $m_{\beta\beta}$ and is the primary challenge for interpreting experimental half-lives.

If neutrino masses obey the so-called inverted ordering, where a single mass eigenstate is substantially lighter than the other two --- a question that should be answered by upcoming oscillation experiments such as DUNE \cite{dune} and Hyper-Kamiokande \cite{hyper-k} --- then measurements of neutrino mass differences and mixing angles make it possible to bound $m_{\beta\beta}$ from below.  This raises the tantalising possibility that any result, positive or negative, from next-generation $0\nu\beta\beta$ experiments, will provide a definitive answer to whether the Majorana mass term in Eq.~(\ref{majorana-mass}) is present in the Lagrangian of nature.  However, this relies on improving theoretical constraints on the relevant nuclear matrix elements in order to properly constrain $m_{\beta\beta}$ from $0\nu\beta\beta$ searches.

Nuclear effective field theory (EFT) offers a solution to this problem by providing a model-independent solution of nuclear physics.  Treating nucleons as the effective degrees of freedom of the problem and integrating out high-energy quark and gluon interactions gives rise to a set of couplings among nucleons (and, depending on the energy range being examined, possibly pions as well).  Quark-level interactions with electroweak fields give rise to single- and multi-nucleon interactions with axial and vector external currents.

Such couplings can be computed from lattice QCD or fit to experimental data from light nuclei, such as deuterium and tritium.  Nuclear EFT, combined with various many-body methods, then allows predictions of masses and single-$\beta$ decay rates in larger nuclei in reasonable agreement (often 10--20\%) with direct experimental measurements \cite{pionless-eft-calcium, chiral-eft-tin}.

\begin{figure}[h]
  \centering
  \includegraphics[align=c,width=0.49\textwidth]{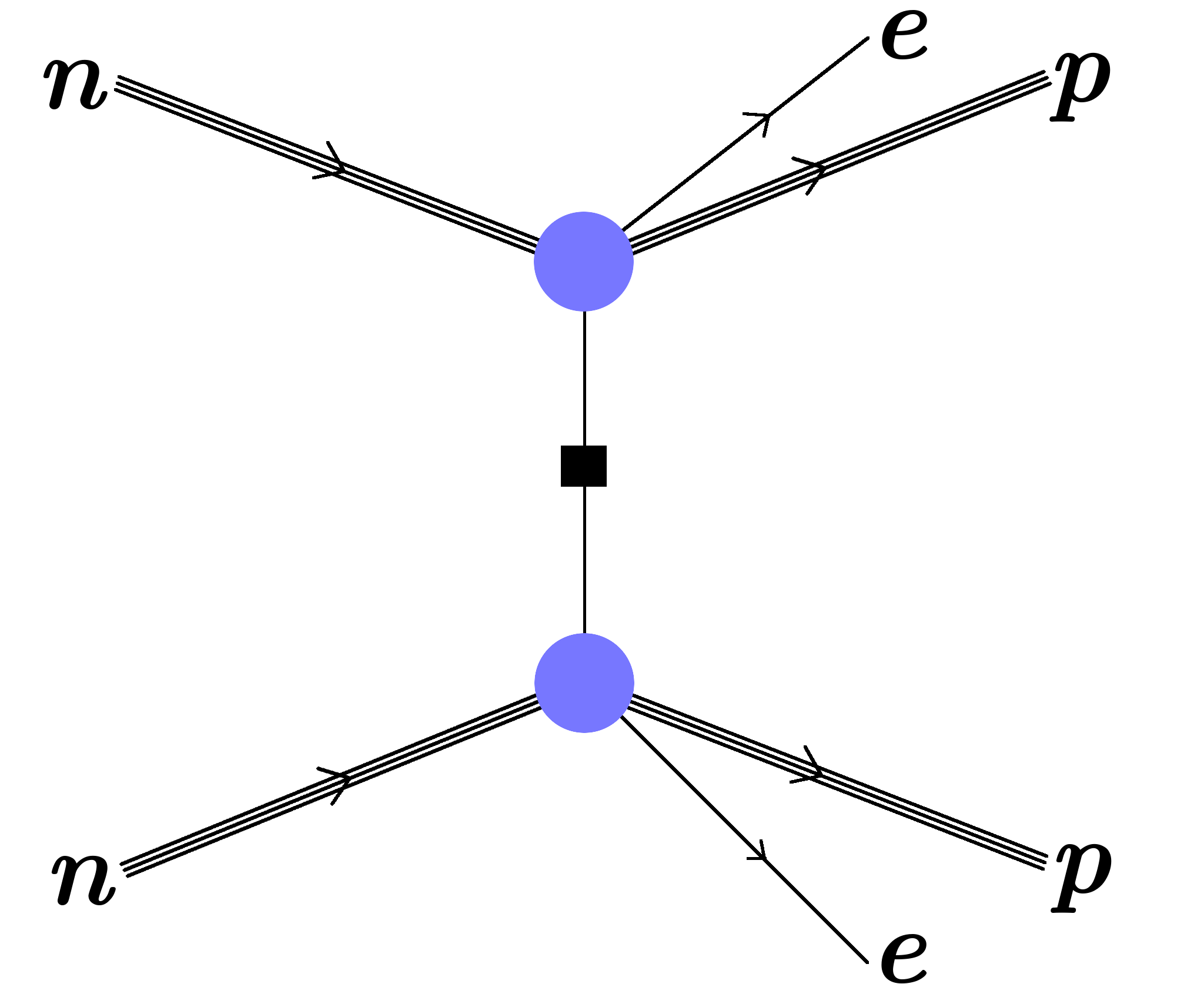}
  \includegraphics[align=c,width=0.49\textwidth]{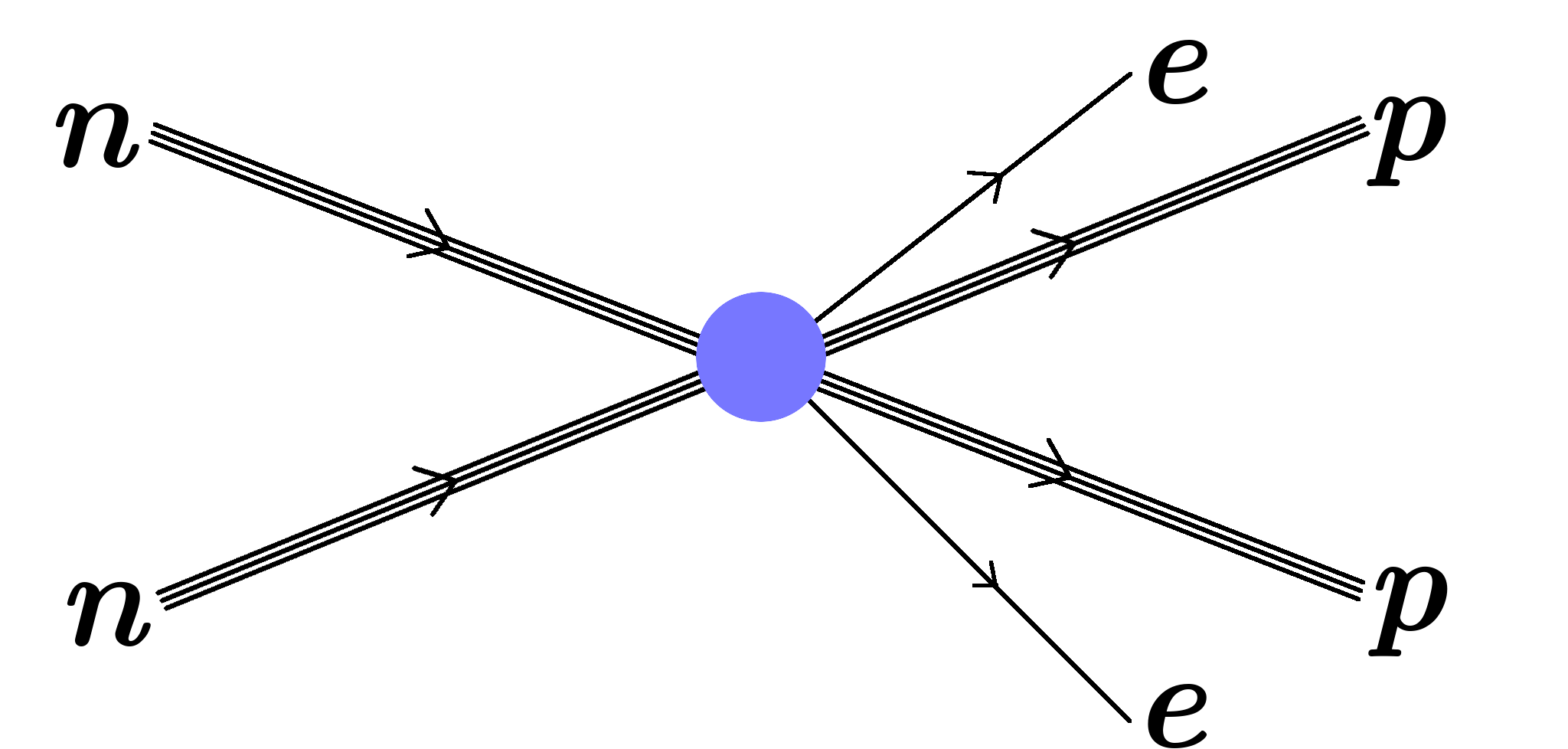}
  \caption{The diagrams contributing to $0\nu\beta\beta$ in nuclear EFT.  (Left) $0\nu\beta\beta$ can proceed via two widely separated electroweak current insertions exchanging a low-energy Majorana neutrino.  This can be calculated from single-$\beta$ decay rates with no additional EFT parameters.  (Right) There is also a contribution from a four-nucleon contact interaction that arises from exchange of a high-energy neutrino with momentum above the cutoff for the EFT validity.  This contribution is also leading-order, and the interaction strength of this vertex is a free parameter in the EFT that must be determined by matching onto lattice QCD or some model of nuclear interactions.}
  \label{0vbb-diagrams}
\end{figure}

At the level of nuclear EFT, two diagrams contribute to neutrinoless double-beta decay, as shown in Fig.~\ref{0vbb-diagrams}.  The process can occur via two weak current insertions with a soft neutrino propagating between them.  However, to handle reactions where the neutrino energy is large relative to the cutoff scale of the EFT (typically taken as $m_\pi$ for pionless EFT), then a four-nucleon contact interaction must be added to the EFT, and this term must be promoted to leading order in the EFT power counting \cite{cirigliano-1, cirigliano-2}.

The first diagram in the EFT can be computed from nucleon axial and vector couplings, which can be fit to single-$\beta$ decay rates.  The second diagram, however, is unique to $0\nu\beta\beta$, and its coefficient $g_{NN}^\nu$ cannot be determined experimentally since $0\nu\beta\beta$ decay has never been observed.  The best estimates on its magnitude in the literature come from dispersive relations \cite{cirigliano-3} that generalize the Cottingham formula for electromagnetic corrections to nuclear masses \cite{cottingham} and from the large-$N_c$ limit of QCD \cite{richardson-1, richardson-2}.  These results are in good agreement with each other within about 30--40\% uncertainties but require model assumptions.  Work is ongoing to refine these calculations \cite{van-goffrier}, but to date there is no fully model-independent calculation of $g_{NN}^\nu$.  Lattice QCD offers the possibility of such a computation using a small nuclear system (namely, the $nn \rightarrow pp$ transition), and this work will present progress that has been made toward this goal.  Sect.~\ref{previous-work} will survey the double-beta computations performed in lattice QCD, Sect.~\ref{future-work} will detail future challenges and sketch a roadmap of where progress will hopefully be made in coming years, and Sect.~\ref{conclusion} will conclude.

\section{Lattice QCD Calculations}
\label{previous-work}
The lattice QCD community has risen to the challenge of computing double-beta matrix elements in simple systems toward the goal of extracting $g_{NN}^\nu$.  Pionic systems, while being useful constraints on EFT predictions in their own right, are cleaner systems in order to perform full analyses of all statistical and systematic uncertainties involved in a physical-point calculation.  Multi-nucleon systems, even with $A=2$, are more challenging, so only single-ensemble studies have been performed to date.  While an extraction of $g_{NN}^\nu$ from lattice QCD is not yet possible, these studies have laid the groundwork for future calculations that will be able to compute this EFT coefficient.

\subsection{\texorpdfstring{$0\nu\beta\beta$ in $\pi^- \rightarrow \pi^+$}{0vbb in pi- -> pi+}}
\label{pion-0vbb}
Pion interactions can be explicitly included in nuclear calculations using chiral effective field theory ($\chi$EFT), improving the accuracy of the theory and extending the range of energies over which the EFT is valid at the cost of introducing additional low-energy constants.  Of particular interest is the constant $g_{\pi\pi}^\nu$, the analogue of $g_{NN}^\nu$ for the $\pi^- \rightarrow \pi^+ ee$ transition.

\begin{figure}[h]
  \centering
  \includegraphics[width=0.49\textwidth]{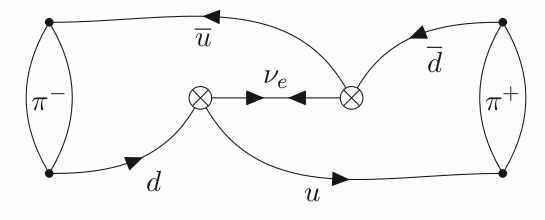}
  \includegraphics[width=0.49\textwidth]{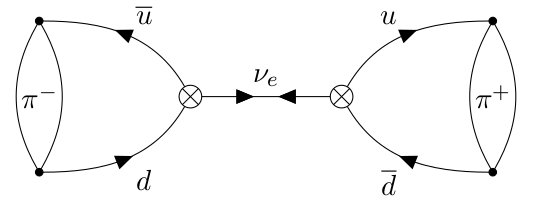}
  \caption{The two Feynman diagrams contributing to the neutrinoless double-beta decay reaction $\pi^- \rightarrow \pi^+ ee$.  The first diagram is fully connected, but the second has no connected quark propagators connecting the source and sink, which are coupled only by the neutrino.  Figure taken from Ref.~\cite{detmold-murphy}.}
  \label{0vbb-pi-diagrams}
\end{figure}

In lattice QCD, the correlation function corresponding to this transition is \cite{detmold-murphy, tuo}
\begin{equation}
  C_{\pi^- \rightarrow \pi^+}(t_+, t, t_-) = \sum_{\mathbf{x}, \mathbf{y}} \int \frac{d^4 q}{(2\pi)^4} \frac{e^{i q \cdot (x-y)}}{q^2} \langle \mathcal{O}_{\pi^+} (t_+) J_\mu(x) J_\mu (y) \mathcal{O}^\dagger_{\pi^-}(t_-) \rangle \, ,
  \label{0vbb-pi-correlator}
\end{equation}
where $O_{\pi} (t)$ is the pion interpolating operator at timeslice $t$, $J_\mu (x) = [\bar u (1 - \gamma_5) \gamma_\mu d](x)$ is the weak current insertion at spacetime position $x$, and $t = x - y$.  This gives rise to two diagram topologies --- one connected and one disconnected --- shown in Fig.~\ref{0vbb-pi-diagrams}.  In both calculations of this quantity, the correlation functions were computed by constructing wall sources and sinks at all timeslices and contracting them at the operator insertion points $x$ and $y$.  

At a fixed set of times, this na\"ively costs $O(V^2) = O(L^6)$ to perform the double sum over spatial positions.  A key insight in Ref.~\cite{detmold-murphy-1} is that the double sum is a convolution, which the Fourier convolution theorem allows one to write as a product of fast Fourier transforms \cite{fft}, decreasing the contraction cost to the more tractable $O(L^3 \log L)$.  One can then compute $C_{\pi^- \rightarrow \pi^+}$ as a function of the various separation times between source, sink, and operators.

The second diagram topology in Fig.~\ref{0vbb-pi-diagrams} contains an intermediate state (the QCD vacuum and a massless neutrino) with lower energy than the initial and final states, so its contribution to a Euclidean-space correlation function grows exponentially rather than decaying with operator separation time $t$.  This is an artifact of Euclidean time rather than a physically diverging amplitude, so this must be corrected before a meaningful matrix element can be extracted from the correlation function.  In this case, the divergent contribution can be written in terms of the pion-to-vacuum matrix element and computed in terms of the pion decay constant $f_\pi$, allowing this to be removed prior to integration over $t$.  The removal of this contribution can then be corrected after the conversion to Minkowski space.

The computations in Refs.~\cite{tuo, detmold-murphy} were both performed with domain wall quarks and a range of lattice spacings and volumes, as well as a quark mass tuned to the physical point (Ref.~\cite{tuo}) or over a wide enough range of pion masses to allow for a physical-point extrapolation (Ref.~\cite{detmold-murphy}).  Chiral perturbation theory relates the computed matrix element $M^{0\nu}$ to the EFT coefficient $g_\nu^{\pi\pi}$ via
\begin{equation}
  \frac{M^{0\nu}}{M^{0\nu}_{(0)}} = 1 + \frac{m_\pi^2}{8\pi^2 f_\pi^2} \left( 3 \log \frac{\mu^2}{m_\pi^2} + 6 + \frac{5}{6}g_\nu^{\pi\pi}(\mu) \right)
  \label{chpt-gpipi}
\end{equation}
where $M^{0\nu}_{(0)}$ is the analytically calculable contribution from the second diagram in Fig.~\ref{0vbb-diagrams}, $f_\pi \approx 130$ MeV is the pion decay constant, and $\mu$ is the renormalization scale.  At a scale of $\mu = m_\rho = 770$ MeV, $g_\nu^{\pi\pi}(\mu)$ was computed as $-10.9(8)$ in Ref.~\cite{tuo} and $-10.8(5)$ in Ref.~\cite{detmold-murphy}, where the errors include all statistical and systematic uncertainties.

\subsection{Dimension-Nine Matrix Elements}
The preceding sections have focused on the minimal way to include a Majorana neutrino mass into the Standard Model, namely, by combining Eq.~(\ref{majorana-mass}) with Standard Model left-handed electroweak interactions.  While Eq.~(\ref{majorana-mass}) is universal and appears in any theory in which the neutrino is Majorana, some beyond-the-Standard-Model theories postulate additional lepton-number-violating interactions above the electroweak scale, which can be integrated out to give four-quark dimension-nine interaction vertices of the form
\begin{equation}
  \mathcal{O} = (\bar d \Gamma_i u) (\bar d \Gamma_j u) (\bar e \Gamma_k e^C)
  \label{4-quark-operator}
\end{equation}
at the electroweak scale and below.  Note that, unlike the EFT contact operator on the right-hand side of Fig.~\ref{0vbb-diagrams}, which is short-distance at nuclear scales but long-distance at quark scales, the contact interaction in Eq.~(\ref{4-quark-operator}) is short-distance even at the quark level.

Using various Fierz relations, one can show that there are nine possible operators of the general form in Eq.~(\ref{4-quark-operator}), of which five are scalar operators (canonically indexed $\mathcal{O}_1, \mathcal{O}_2, \mathcal{O}_3, \mathcal{O}_1', \mathcal{O}_2'$) and four are vector operators ($\mathcal{V}_1, \mathcal{V}_2, \mathcal{V}_3, \mathcal{V}_4$) depending on whether $\Gamma_k$ carries a free Lorentz index \cite{heavy-neutrino-operators, heavy-neutrino-operators-2, Cirigliano_2020}.  Predictions from general BSM models can be written as linear combinations of these nine operators.

In the $\pi^- \rightarrow \pi^+ ee$ transition, only the scalar operators contribute meaningfully (with the vector operators suppressed by $O(m_e / m_\pi) < 10^{-2}$).  Weinberg power counting \cite{chiral-eft-1, chiral-eft-2, chiral-eft-3} predicts that $\pi$ exchange, and therefore the $\pi^- \rightarrow \pi^+ ee$ subprocess, dominates $0\nu\beta\beta$ in nuclear systems as well \cite{Cirigliano_2018}, although this prediction is less certain due to theoretical difficulties with Weinberg power counting.

The matrix elements of these five scalar operators have been computed in the $\pi^- \rightarrow \pi^+ ee$ transition by both the CalLat \cite{nicholson} and NPLQCD \cite{nplqcd-short-distance} collaborations.  As with their long-distance counterpart, these matrix elements have been extrapolated to the physical point with all lattice artifacts controlled.  The corresponding values are tabulated in Tab.~\ref{short-distance-results}.

\begin{table}
  \centering
  \begin{tabular}{|c|c|c|} \hline
    $\langle \pi^+ | \mathcal{O} | \pi^- \rangle$ (GeV$^4$) & CalLat & NPLQCD \\ \hline
    $\mathcal{O}_1$ & $-1.91(13) \times 10^{-2}$ & $-1.27(16) \times 10^{-2}$ \\
    $\mathcal{O}_2$ & $-3.68(31) \times 10^{-2}$ & $-2.45(22) \times 10^{-2}$ \\
    $\mathcal{O}_3$ & $\phantom{-}1.85(10) \times 10^{-4}$ & $\phantom{-}0.87(8)~ \times 10^{-4}$ \\
    $\mathcal{O}_{1'}$ & $-7.22(49) \times 10^{-2}$ & $-5.35(48) \times 10^{-2}$ \\
    $\mathcal{O}_{2'}$ & $\phantom{-}1.16(10) \times 10^{-2}$ & $\phantom{-}0.76(8)~ \times 10^{-2}$ \\ \hline
  \end{tabular}
  \caption{Dimension-nine (short-distance) matrix elements for the $\pi^- \rightarrow \pi^+ ee$ transition, from the CalLat \cite{nicholson} and NPLQCD \cite{nplqcd-short-distance} collaborations.  Results are listed in the $\overline{\textrm{MS}}$ scheme at a renormalisation scale of $\mu = 3$ GeV.}
  \label{short-distance-results}
\end{table}

Whilst there is some tension between the results, there is qualitative agreement in the magnitudes of the various matrix operators.  In particular, $\mathcal{O}_3$ is smaller than the others by a factor of about $10^2$.  This agrees with predictions from $\chi$PT, which suggest that $\mathcal{O}_3$ is suppressed by a factor of $m_\pi^2 / (4\pi f_\pi)^2$.

\subsection{Neutrinoful Double-Beta Decay}
Nuclear systems pose several challenges avoided by the pionic calculations.  First, baryons --- and especially nuclei with multiple baryons --- suffer from exponentially bad signal-to-noise degredation at large source-sink separations.  Additionally, the number of contractions grows factorially with the number of quarks in the system, and even for two-baryon systems, the contraction cost can rival that of propagator inversion.

The first problem can be mitigated, although not altogether avoided, by working with larger-than-physical quark masses.  At $m_\pi \approx 800$ MeV, it is mild enough to have enabled spectroscopy calculations of light nuclei for over a decade \cite{light-nuclei-nplqcd, pp-fusion-lattice, nplqcd-2, nplqcd-4, nplqcd-variational, nucleon-scattering, callat-variational, mainz-variational}.  While this unphysical mass means that the result cannot be directly compared to nature, it is a useful starting point for a campaign of computing double-beta decay matrix elements that will eventually be extrapolated to physical quark masses.

The problem of nuclear contractions is simplified if one instead studies
the neutrinoful double-beta decay reaction in Eq.~(\ref{neutrinoful-double-beta-decay}).
Without the Majorana neutrino propagating between the two electroweak currents, these two current insertions decouple and can be computed more easily.

In particular, the background field method \cite{background-field-1, background-field-2, background-field-3} can be used to compute electroweak matrix elements.  Propagators are computed both in pure QCD and also in the presence of a background field with strength proportional to $\lambda$.  The derivative of the two-point correlation function with respect to $\lambda$ is proportional to the forward matrix element corresponding to the background field (e.g.~the electric or axial charge, depending on the specific field used).

This method can be generalized to multiple current insertions, and, in particular, to $2\nu\beta\beta$ matrix elements.  Here, for a second-order weak process, one seeks the second derivative with respect to $\lambda$.  
At the cost of computing propagators at several values of $\lambda$ to compute the second derivative numerically, this reduces the problem of computing four-point correlation functions to that of two-point correlation functions.  In particular, one does not need to construct propagators coming from the sink (as was done in the pionic calculations), nor does one need to contract propagators at various operator separations and momenta.
As a result, the $2\nu\beta\beta$ matrix element can be computed to high (few-percent) statistical precision.

As with its neutrinoless counterpart, the nuclear EFT includes contributions from two well-separated electroweak insertions as well as a contact operator with a low-energy constant $\mathbb{H}_{2,S}$ that is unknown \emph{a priori} \cite{2vbb-decay-lattice}:
\begin{equation}
  M^{2\nu} = -\frac{|M_{pp\rightarrow d}|^2}{\Delta} + \frac{M g_A^2}{4 \gamma_s^2} - \mathbb{H}_{2,S}
  \label{neutrinoful-double-beta-eft}
\end{equation}
where $|M_{pp \rightarrow d}|$ is the single-$\beta$ decay matrix element, $\Delta = E_{nn} - E_d$ is the deuteron-dineutron mass splitting, $g_A$ is the single-nucleon axial charge, and $\gamma_s$ is the dineutron binding momentum.
Unlike in $0\nu\beta\beta$, however, the contact operator in $2\nu\beta\beta$ is next to leading order in the EFT power counting.
The high-precision calculation of $M^{2\nu}$ in Ref.~\cite{2vbb-decay-lattice} resolved the full matrix element from the leading-order EFT prediction, finding $\mathbb{H}_{2,S}$ to be about a 5\% correction to the full matrix element at the lattice spacing and pion mass considered in that work.

\subsection{\texorpdfstring{$0\nu\beta\beta$ for $nn \rightarrow pp$}{0vbb for nn->pp}}
\label{nnpp}
The success of pionic calculations and of $2\nu\beta\beta$ in nuclear systems motivates the more computationally demanding calculation of $0\nu\beta\beta$ in the $nn \rightarrow pp ee$ reaction.  As in the $2\nu\beta\beta$ calculation in the previous section, the signal-to-noise problem favours performing this computation at unphysically heavy quark masses.

Unlike in the $2\nu\beta\beta$ calculation where the background field method could be utilised, the four-point correlation function for $0\nu\beta\beta$ must be computed explicitly.  Two stategies have emerged for doing so:
\begin{enumerate}
  \item Compute propagators coming from both source and sink and contract them at the operator insertion points, as was done in the $\pi^- \rightarrow \pi^+ ee$ calculations in Sect.~\ref{pion-0vbb}.  However, the wall source/sink combination works substantially better for pions than for nucleons, and volume-averaged point (or smeared) sinks commonly used in two-point correlation functions would here require a separate inversion from every sink point.  This necessitates some cost-reduction strategy at the sink, either sparsening \cite{sparsening} or distillation \cite{distillation-1, distillation-2, distillation-3}.
  \item Compute sequential propagators through the operator insertion points.  The primary difficulty here is that the intermediate neutrino energy must be integrated over, na\"ively requiring separate sequential inversions for every possible choice of $\mathbf{p}_\nu$.  A stochastic propagator for the neutrino dramatically reduces this number of inversions at the cost of adding additional noise to the correlation function.
\end{enumerate}

In Ref.~\cite{0vbb-paper}, the first option was chosen, with a sparse grid of $4^3$ propagators computed at the sink.  Ideally, one would use an identical sparse grid of propagators at the source to obtain a symmetric correlation function, as has been done in recent nuclear spectroscopy calculations \cite{nplqcd-variational, callat-variational, mainz-variational}.  However, the computational cost of computing 4-point functions substantially exceeds that of the 2-point correlation functions used in spectroscopy: propagators are needed that originate over a range of source and sink timeslices, increasing inversion costs by an order of magnitude or more, and contractions must be computed over all combinations of source, sink, and operator times, of which there can be several hundred.  As a cost-saving measure, wall sources were used, 
requiring only one additional inversion when changing the source-sink separation.  Furthermore, unlike the bi-local operators used in spectroscopy calculations, the computationally cheaper local hexaquark operator was used at the sink.

In spite of these simplifications, performing the nuclear contractions efficiently is a challenging computational problem.  For fixed operator and sink positions, the number of quark contractions required for the $nn \rightarrow pp ee$ matrix element is roughly $N_c!^4 N_u! N_d! \approx 10^6$, and na\"ively this cost must be repeated at each of the operator positions, contributing an additional factor of $V^2$.  Additional multiplicative factors arise from varying the sink spatial position and the temporal positions of operators and source.

The Fourier convolution trick used in the pionic calculations reduces the cost of the operator sum to $V \log V$ times the number of nuclear contractions.  One might attempt to further reduce this cost by sparsening at the operator positions as is done at the sink.  However, sparsening effectively truncates the spectrum in momentum space, leading to a distortion of high-momentum modes.  At source and sink, such high-momentum modes correspond to excited states that are removed by time evolution, but in four-point functions where the neutrino momentum must be integrated over, operator sparsening gives an incorrect result (shown in preliminary testing to be wrong by a factor of several).

However, it is possible to decouple the volume sum for operator insertions from the $O(10^6)$ nuclear contractions.  If $S_{cb}^{\zeta\beta}(x | x_i)$ is the propagator from source to operator, $S_{ae}^{\alpha\delta}(x_f | x)$ is the analogous propagator originating at the sink, and $\left( J^\mu \right)^{\delta\zeta}$ is the weak current insertion, then one can compute sequential propagators
\begin{equation}
  S_{ab}^{\alpha\beta, \mu} (x) = S_{ae}^{\alpha\delta}(x_f | x) \left( J^\mu \right)^{\delta\zeta} S_{eb}^{\zeta\beta}(x | x_i)
  \label{0vbb-seqprop}
\end{equation}
and convolve the two Fourier-transformed sequential propagators for the two electroweak insertions with the neutrino propagator $D(x-y)$ to obtain
\begin{align}
  T_{abcd}^{\alpha\beta\gamma\delta} = \frac{a^3}{L^3} \sum_{\mathbf{p}} \mathcal{F}[S^{\alpha\beta,\mu}_{ab}](\mathbf{p}; t_x) \mathcal{F}[D](p; t_x-t_y) \mathcal{F}[S_{cd}^{\gamma\delta, \mu}](-\mathbf{p}, t_y)
  \label{four-quark-tensor}
\end{align}
which has four free spin-color indices but is independent of the nuclear contractions at source and sink.  If all quarks in the source and sink interpolating operators are projected to positive parity, e.g.~if $\mathcal{O}_p^\delta = \varepsilon_{abc} [u_a^\alpha (P_+ C\gamma_5)^{\alpha\beta} d_b^\beta] (P_+)^{\delta\gamma} u_c^\gamma$, then the cost of computing $T_{abcd}^{\alpha\beta\gamma\delta}$ is $O(6^4 V \log V + 10^6)$, substantially faster than $O(10^6 V \log V)$.

A final challenge is the power-law fall-off of the neutrino propagator $D(x-y) \sim t^{-2}$ at large operator separations, which is much slower than the exponential signal degredation at large $t$.  This can be ameliorated by considering a zero-mode-subtracted version of the propagator \cite{zohreh-nu-prop}
\begin{equation}
  D(x, y) = \frac{1}{2L^3} \sum_{\mathbf{q} \in \frac{2\pi}{L}\mathbb{Z}^3\\\left\{ 0 \right\}}^{|\mathbf{q}|\leq \pi/a} \frac{1}{|\mathbf{q}|} e^{i\mathbf{q} \cdot (\mathbf{x}-\mathbf{y})} e^{-|\mathbf{q}| |t|}
  \label{nu-prop-subtracted}
\end{equation}
that decays exponentially in $t$ in finite volume.  The lattice QCD matrix element is then matched to an EFT computed with the same neutrino propagator in order to extract $g_{NN}^\nu$.  

Using this propagator, the matrix element $\mathcal{A}^{0\nu}$ (proportional to $M^{0\nu}$ in Eq.~(\ref{decay-relation}) above) is computed by first extrapolating the ratio $R(t_\text{snk}, t, t_\text{src})$ of 4-point to 2-point correlation functions to large source-operator and sink-operator separations $t_\text{src}$ and $t_\text{snk}$ in order to remove excited state contamination to obtain $R(t) = \lim_{\substack{t_\text{src} \rightarrow \infty \\ t_\text{snk} \rightarrow \infty}} R(t_\text{snk}, t, t_\text{src})$.  Then integrating over operator separation times $t$ gives
\begin{equation}
  \mathcal{A}^{0\nu} = 2 E_0 \int_{-\infty}^\infty dt \, R(t) \, .
  \label{0vbb-matrix-element-integral}
\end{equation}
In practice, the large-$t$ behaviour of $R(t)$ is poorly constrained due to the signal-to-noise problem, so $R(t)$ is fit to an exponential in $t$ that can be integrated analytically.  At a fixed lattice spacing and pion mass, Ref.~\cite{0vbb-paper} used this procedure to determine $\mathcal{A}^{0\nu}$ to about 20\% precision.

More recently, a preliminary study has been undertaken using stochastic neutrino propagators in order to avoid the additional cost of having propagators sourced at both source and sink \cite{zi-yu-presentation}.  After computing these stochastic sequential propagators through the operator insertion points, the nuclear contractions are done in a similar way to standard two-point correlation functions.  In particular, a setup with symmetric bi-local interpolating operators at source and sink becomes feasible, reducing excited state contamination at the cost of the additional noise from the stochastic neutrino propagator.

While this work is still unpublished, preliminary results have been presented at this conference, indicating reasonable ($\lesssim 10\%$) statistical noise over moderate operator separations.  Perhaps more importantly, this study was performed using quark masses corresponding to $m_\pi = 432$ MeV, about half the pion mass used in the previous study.  This represents important progress toward physical-point simulations.

\section{Remaining Challenges and Future Work}
\label{future-work}
In order to make contact with experimental results, the lattice QCD community must be able to provide $g_{NN}^\nu$ at the physical point with all uncertainties fully controlled.  Whilst much work has already been completed toward this goal, much work remains to be done in the coming years, ideally in time to interpret results from next-generation $0\nu\beta\beta$ experiments.

\subsection{EFT Matching}
From an EFT perspective, the matrix element $\mathcal{A}^{0\nu}$ for the $0\nu\beta\beta$ reaction computed in the previous section originates as the sum of the long-distance and contact-term diagrams shown in Fig.~\ref{0vbb-diagrams}.  More concretely, one can write \cite{zohreh-nu-prop}
\begin{equation}
  \frac{\mathcal{A}^{0\nu}}{2m_{nn}} \frac{1}{\mathcal{R}(E) \mathcal{M}(E)} = (1+3g_A^2) (J^\infty + \delta J^V) - \frac{m_n^2}{8\pi^2} \tilde g_\nu^{NN}
  \label{eft-matching-0vbb}
\end{equation}
where $\mathcal{M}(E)$ is the $NN$ scattering amplitude, $\mathcal{R}(E)$ is the Lellouch-L\"uscher residue, $J^\infty$ is the contribution of the first diagram in Fig.~\ref{0vbb-diagrams}, $g_A$ is the axial charge of the nucleon, $\delta J^V$ is a finite-volume correction \cite{zohreh-nu-prop, davoudi-2}, and $\tilde g_\nu^{NN}$ is $g_\nu^{NN}$ rescaled by normalisation factors.  The functional forms of these functions are well known in the literature, but they have free parameters that must be determined from $NN$ interactions.  For example, the effective-range expansion for the $NN$ scattering amplitude,
\begin{equation}
  \mathcal{M}(E) = -\frac{4\pi}{m_N} \frac{1}{1/a - rp^2/2 + ip}
  \label{effective-range-expansion}
\end{equation}
depends on the scattering length $a$, the effective range $r$, and the finite-volume energy shift $E = p^2 / 2m_N$.

In lattice QCD, these values are also calculable in principle by precisely determining the spectrum of two nucleons within a finite volume and computing the energy shifts of the spectrum relative to non-interacting energy levels.  An important challenge in such a calculation is difficult-to-control systematic effects arising from excited-state contamination.  In particular, computations of the ground state energy of the ensemble used in the $0\nu\beta\beta$ calculation in Ref.~\cite{zohreh-nu-prop, davoudi-2}
differ by several standard deviations depending on whether one uses asymmetric correlation functions (which find a deep bound state dineutron \cite{light-nuclei-nplqcd, pp-fusion-lattice, nplqcd-2, nplqcd-4, nucleon-scattering}) or variational methods using symmetric correlation functions (which do not provide evidence of a bound state \cite{callat-variational, mainz-variational, nplqcd-variational}).  The difference between the values of $\tilde g_{NN}^\nu$ computed using inputs from these two methods far exceeds the uncertainty arising from the $0\nu\beta\beta$ matrix element computed in Ref.~\cite{0vbb-paper}, and assessing the validity of the various spectroscopic calculations is beyond the scope of this work.

A computation of $\mathcal{A}^{0\nu}$ directly at the physical point would not suffer from this difficulty, as the scattering length and effective range are well known experimentally (to few-percent precision), and the finite-volume energy shift can be computed from these inputs.  However, if $\mathcal{A}^{0\nu}$ is computed at various unphysical quark masses such that $\tilde g_{NN}^\nu$ can be extracted and extrapolated to the physical point, then corresponding spectroscopic calculations on the various ensembles are needed to perform the EFT matching.

\subsection{Towards Physical Point Calculations}
At lighter quark masses, lattice QCD calculations become more challenging for several reasons, most saliently higher costs for propagator inversions and a worsened signal-to-noise problem in baryonic systems.  Much work has been done to ameliorate the first of these challenges --- in particular, the use of algebraic multigrid preconditioners in solvers \cite{multigrid-1, multigrid-2, multigrid-3} --- but the second remains an open challenge.  For two-baryon systems, this is especially challenging, with the signal-to-noise ratio expected to fall off asymptotically twice as fast in time as in the single-baryon case \cite{parisi, lepage}.  Four-point correlation functions are more challenging still, as computing the matrix element requires access to a range of source-operator, operator-operator, and operator-sink separations before the signal is lost to noise.

The range of operator-operator separations over which one must scan is a physically relevant length scale dictated by the rate of the neutrino propagator decay, but separations from the source and sink are only needed to control excited state contamination whose fall-off depends on the choice of interpolating operators.  Thus, the choice of interpolating operators with minimal excited state contamination will become even more important with light quarks.  In particular, the trade-off between computational cost of an interpolating operator and control over its overlap with excitations becomes tilted more heavily toward the latter when near-physical masses are used.

Two-point spectroscopic calculations can be informative in this regard.  Variational analyses, which solve the generalised eigenvalue problem (GEVP) among a class of interpolating operators, can determine the coefficients of the optimal linear combination of these operators that minimises excited state contamination.  This linear combination can then be used in computing four-point functions for $0\nu\beta\beta$, and operators with sufficiently small coefficients in the GEVP can be excluded in order to reduce computational costs.

Recently, variational calculations of the two-nucleon spectrum at near-physical masses have begun using various bi-local interpolating operators for the dineutron and deuteron with different relative momenta.  The precision of such calculations is not yet sufficient to resolve finite-volume energy shifts, but preliminary results show optimism for such resolution with increased statistics, hopefully within timescales of a year or two.
The work in Ref.~\cite{zi-yu-presentation} has shown one method for incorporating bi-local interpolating operators into $0\nu\beta\beta$ calculations.  Future work will hopefully explore whether the method of performing $0\nu\beta\beta$ contractions at the operator insertion points can be sufficiently optimised so as to be compatible with these bi-local interpolating operators as well.

\section{Conclusion}
\label{conclusion}
The scientific community is at an exciting juncture in the study of neutrinoless double-beta decay.  Next-generation experiments with the potential to probe the entire inverted hierarchy of neutrino masses will come online within the next decade and are scheduled to release results within a decade following that, if not earlier.  Within such a timescale, it is incumbent on the theory community to resolve the large uncertainties in nuclear matrix elements that currently limit the interpretability of experimental results.

Computing such matrix elements with controlled uncertainties will take a concerted effort of the nuclear theory community, including advances in nuclear many-body calculations that will use inputs from nuclear effective field theory to make predictions about the medium and large nuclei that are experimentally relevant.  The main role of lattice QCD practitioners is to provide the necessary low-energy constants of the EFT, in particular the coefficient of the four-nucleon contact interaction $g_{NN}^\nu$, in time to be useful as an input.

Physical-point nuclear spectroscopy calculations are currently ongoing and optimistically will observe a loosely bound deuteron within the next year or two.  In parallel to this, it would be beneficial to perform a comparative study of the relative cost and signal quality of the various possible methods for computing the necessary four-point correlation functions for $0\nu\beta\beta$.  With these prerequisites, the first $0\nu\beta\beta$ calculation at near-physical quark masses is optimistically possible within a three-to-five-year timescale, with a controlled continuum extrapolation a couple years after that.  Combined with advances in many-body calculations over a similar timescale, nuclear theory could proved these matrix elements on an experimentally relevant timescale.

\acknowledgments
The author would like to thank Zohreh Davoudi, William Detmold, Xu Feng, Zhenghao Fu, Daniel Hackett, Marc Illa, William Jay, Emanuele Merghetti, David Murphy, Amy Nicholson, Patrick Oare, Assumpta Parre\~no, Robert Perry, Phiala E. Shanahan, Ruth Van de Water, Michael L. Wagman, Andr\'e Walker-Loud, and Zi-Yu Wang for helpful discussions.
This manuscript has been authored by Fermi Forward Discovery Group, LLC under Contract No. 89243024CSC000002 with the U.S. Department of Energy, Office of Science, Office of High Energy Physics.

\bibliographystyle{JHEP}
\bibliography{refs}

\providecommand{\href}[2]{#2}\begingroup\raggedright\begin{thebibliography}{10}

\bibitem{homestake}
J.N.~Bahcall and R.~Davis, \emph{Solar neutrinos: A scientific puzzle},
  \href{https://doi.org/10.1126/science.191.4224.264}{\emph{Science} {\bfseries
  191} (1976) 264}.

\bibitem{super-kamiokande}
S.~Fukuda et~al., \emph{The {S}uper-{K}amiokande detector},
  \href{https://doi.org/10.1016/S0168-9002(03)00425-X}{\emph{Nucl. Instrum.
  Methods Phys. Res. A} {\bfseries 501} (2003) 418}.

\bibitem{sno}
Q.R.~Ahmad et~al., \emph{Measurement of the rate of $\nu_e + d \rightarrow p +
  p + e^-$ interactions produced by $^8${B} solar neutrinos at the {S}udbury
  {N}eutrino {O}bservatory},
  \href{https://doi.org/10.1103/PhysRevLett.87.071301}{\emph{Phys. Rev. Lett.}
  {\bfseries 87} (2001) 071301}
  [\href{https://arxiv.org/abs/nucl-ex/0106015}{{\ttfamily nucl-ex/0106015}}].

\bibitem{schwartz}
M.D.~Schwartz, \emph{Quantum Field Theory and the Standard Model}, Cambridge
  University Press, Cambridge, United Kingdom, 3rd~ed. (2015).

\bibitem{majorana}
E.~Majorana, \emph{Teoria simmetrica dell'elettrone e del positrone},
  \href{https://doi.org/10.1007/BF02961314}{\emph{Nuovo Cim.} {\bfseries 14}
  (1937) 171}.

\bibitem{jaffe-taylor}
R.L.~Jaffe and W.~Taylor, \emph{The Physics of Energy}, Cambridge University
  Press, New York, 1st~ed. (2018).

\bibitem{mayer-double-beta}
M.~Goeppert-Mayer, \emph{Double beta-disintegration},
  \href{https://doi.org/10.1103/PhysRev.48.512}{\emph{Phys. Rev.} {\bfseries
  48} (1935) 512}.

\bibitem{double-beta-discovery}
M.G.~Inghram and J.H.~Reynolds, \emph{Double beta-decay of {Te}$^{130}$},
  \href{https://doi.org/10.1103/PhysRev.78.822.2}{\emph{Phys. Rev.} {\bfseries
  78} (1950) 822}.

\bibitem{double-beta-direct-observation}
S.R.~Elliott, A.A.~Hahn and M.K.~Moe, \emph{Direct evidence for two-neutrino
  double-beta decay in $^{82}${Se}},
  \href{https://doi.org/10.1103/PhysRev.78.822.2}{\emph{Phys. Rev. Lett.}
  {\bfseries 59} (1987) 2020}.

\bibitem{double-beta-status-prospects}
M.J.~Dolinski, A.W.P.~Poon and W.~Rodejohann, \emph{Neutrinoless double-beta
  decay: Status and prospects},
  \href{https://doi.org/10.1146/annurev-nucl-101918-023407}{\emph{Ann. Rev.
  Nucl. Part. Sci.} {\bfseries 69} (2019) 219}
  [\href{https://arxiv.org/abs/1902.04097}{{\ttfamily 1902.04097}}].

\bibitem{jaffe-taylor-source}
J.~Lilley, \emph{Nuclear Physics: Principles and Application}, Wiley, Hoboken,
  NJ (2001).

\bibitem{gerda}
M.~Agostini et~al., \emph{Final results of {GERDA} on the search for
  neutrinoless double-$\beta$ decay},
  \href{https://doi.org/10.1103/PhysRevLett.125.252502}{\emph{Phys. Rev. Lett.}
  {\bfseries 125} (2020) 252502}
  [\href{https://arxiv.org/abs/2009.06079}{{\ttfamily 2009.06079}}].

\bibitem{cuore}
A.~Giachero et~al., \emph{New results from the {CUORE} experiment},
  \href{https://arxiv.org/abs/2011.09295}{{\ttfamily 2011.09295}}.

\bibitem{kamland-zen}
S.~Abe et~al., \emph{Search for the {M}ajorana nature of neutrinos in the
  inverted mass ordering region with {KamLAND-Zen}},
  \href{https://doi.org/10.1103/PhysRevLett.130.051801}{\emph{Phys. Rev. Lett.}
  {\bfseries 130} (2023) 051801}
  [\href{https://arxiv.org/abs/2203.02139}{{\ttfamily 2203.02139}}].

\bibitem{kamland-zen-updated}
S.~Abe et~al., \emph{Search for {M}ajorana neutrinos with the complete
  {KamLAND-Zen} dataset},  \href{https://arxiv.org/abs/2406.11438}{{\ttfamily
  2406.11438}}.

\bibitem{long-range-plan}
G.E.~Dodge, \emph{The {U.S.} nuclear science long range plan},
  \href{https://doi.org/10.1080/10619127.2024.2303306}{\emph{Nucl. Phys. News}
  {\bfseries 34} (2024) 3}.

\bibitem{pontecorvo}
B.~Pontecorvo, \emph{{Inverse beta processes and nonconservation of lepton
  charge}}, {\emph{Zh. Eksp. Teor. Fiz.} {\bfseries 34} (1957) 247}.

\bibitem{mns}
Z.~Maki, M.~Nakagawa and S.~Sakata, \emph{Remarks on the unified model of
  elementary particles}, \href{https://doi.org/10.1143/PTP.28.870}{\emph{Prog.
  Theor. Phys.} {\bfseries 28} (1962) 870}.

\bibitem{dune}
B.~Abi et~al., \emph{{D}eep {U}nderground {N}eutrino {E}xperiment ({DUNE}), far
  detector technical design report, volume {II}: {DUNE} physics},
  \href{https://arxiv.org/abs/2002.03005}{{\ttfamily 2002.03005}}.

\bibitem{hyper-k}
K.~Abe et~al., \emph{{H}yper-{K}amiokande design report},
  \href{https://arxiv.org/abs/1805.04163}{{\ttfamily 1805.04163}}.

\bibitem{nuclear-matrix-elements-plot}
M.~Agostini, G.~Benato, J.A.~Detwiler, J.~Menéndez and F.~Vissani,
  \emph{Toward the discovery of matter creation with neutrinoless $\beta\beta$
  decay}, \href{https://doi.org/10.1103/revmodphys.95.025002}{\emph{Rev. Mod.
  Phys.} {\bfseries 95} (2023) 025002}
  [\href{https://arxiv.org/abs/2202.01787}{{\ttfamily 2202.01787}}].

\bibitem{pionless-eft-calcium}
A.~Bansal, S.~Binder, A.~Ekstr{\"o}m, G.~Hagen, G.R.~Jansen and T.~Papenbrock,
  \emph{Pion-less effective field theory for atomic nuclei and lattice nuclei},
  \href{https://doi.org/10.1103/PhysRevC.98.054301}{\emph{Phys. Rev. C}
  {\bfseries 98} (2018) 054301}
  [\href{https://arxiv.org/abs/1712.10246}{{\ttfamily 1712.10246}}].

\bibitem{chiral-eft-tin}
S.~Binder, A.~Ekstr{\"o}m, G.~Hagen, T.~Papenbrock and K.A.~Wendt,
  \emph{Effective field theory in the harmonic oscillator basis},
  \href{https://doi.org/10.1103/PhysRevC.93.044332}{\emph{Phys. Rev. C}
  {\bfseries 93} (2016) 044332}
  [\href{https://arxiv.org/abs/1512.03802}{{\ttfamily 1512.03802}}].

\bibitem{cirigliano-1}
V.~Cirigliano, W.~Dekens, E.~Mereghetti and A.~Walker-Loud, \emph{Neutrinoless
  double beta decay in effective field theory: the light {M}ajorana neutrino
  exchange mechanism},
  \href{https://doi.org/10.1103/PhysRevC.97.065501}{\emph{Phys. Rev. C}
  {\bfseries 97} (2018) 065501}
  [\href{https://arxiv.org/abs/1710.01729}{{\ttfamily 1710.01729}}].

\bibitem{cirigliano-2}
V.~Cirigliano, W.~Dekens, J.~de~Vries, M.L.~Graesser, E.~Mereghetti, S.~Pastore
  et~al., \emph{A new leading contribution to neutrinoless double-beta decay},
  \href{https://doi.org/10.1103/PhysRevLett.120.202001}{\emph{Phys. Rev. Lett.}
  {\bfseries 120} (2018) 202001}
  [\href{https://arxiv.org/abs/1802.10097}{{\ttfamily 1802.10097}}].

\bibitem{cirigliano-3}
V.~Cirigliano, W.~Dekens, J.~de~Vries, M.~Hoferichter and E.~Mereghetti,
  \emph{Determining the leading-order contact term in neutrinoless double
  $\beta$ decay}, \href{https://doi.org/10.1007/jhep05(2021)289}{\emph{JHEP}
  {\bfseries 05} (2021) 289}
  [\href{https://arxiv.org/abs/2102.03371}{{\ttfamily 2102.03371}}].

\bibitem{cottingham}
W.~Cottingham, \emph{The neutron proton mass difference and electron scattering
  experiments},
  \href{https://doi.org/10.1016/0003-4916(63)90023-X}{\emph{Annals Phys.}
  {\bfseries 25} (1963) 424}.

\bibitem{richardson-1}
T.R.~Richardson, M.R.~Schindler, S.~Pastore and R.P.~Springer,
  \emph{Large-$n_c$ analysis of two-nucleon neutrinoless double-$\beta$ decay
  and charge-independence-breaking contact terms},
  \href{https://doi.org/10.1103/physrevc.103.055501}{\emph{Physical Review C}
  {\bfseries 103} (2021) }.

\bibitem{richardson-2}
T.R.~Richardson, \emph{{Large-$N_c$ constraints for Beyond the Standard Model
  few-nucleon currents in effective field theory}},
  \href{https://doi.org/10.22323/1.413.0098}{\emph{PoS} {\bfseries CD2021}
  (2024) 098}.

\bibitem{van-goffrier}
G.V.~Goffrier, \emph{Nuclear and Particle Physics Aspects of Neutrinoless
  Double-Beta Decay}, Ph.D. thesis, University College London, 2024.

\bibitem{detmold-murphy}
W.~Detmold and D.J.~Murphy, \emph{Neutrinoless double beta decay from lattice
  {QCD}: The long-distance $\pi^- \rightarrow \pi^+ e^- e^-$ amplitude},
  \href{https://arxiv.org/abs/2004.07404}{{\ttfamily 2004.07404}}.

\bibitem{tuo}
X.-Y.~Tuo, X.~Feng and L.-C.~Jin, \emph{Long-distance contributions to
  neutrinoless double beta decay $\pi^- \rightarrow \pi^+ ee$},
  \href{https://doi.org/10.1103/PhysRevD.100.094511}{\emph{Phys. Rev. D}
  {\bfseries 100} (2019) 094511}
  [\href{https://arxiv.org/abs/1909.13525}{{\ttfamily 1909.13525}}].

\bibitem{detmold-murphy-1}
W.~Detmold and D.J.~Murphy, \emph{Nuclear matrix elements for neutrinoless
  double beta decay from lattice {QCD}},
  \href{https://arxiv.org/abs/1811.05554}{{\ttfamily 1811.05554}}.

\bibitem{fft}
J.W.~Cooley and J.W.~Tukey, \emph{An algorithm for the machine calculation of
  complex {F}ourier series},
  \href{https://doi.org/10.1090/S0025-5718-1965-0178586-1}{\emph{Math. Comput.}
  {\bfseries 19} (1965) 297}.

\bibitem{heavy-neutrino-operators}
G.~Pr{\'e}zeau, M.~Ramsey-Musolf and P.~Vogel, \emph{Neutrinoless double-beta
  decay and effective field theory},
  \href{https://doi.org/10.1103/PhysRevD.68.034016}{\emph{Phys. Rev. D}
  {\bfseries 68} (2003) 034016}
  [\href{https://arxiv.org/abs/hep-th/0303205}{{\ttfamily hep-th/0303205}}].

\bibitem{heavy-neutrino-operators-2}
M.L.~Graesser, \emph{An electroweak basis for neutrinoless double $\beta$
  decay}, \href{https://doi.org/10.1007/jhep08(2017)099}{\emph{JHEP} {\bfseries
  08} (2017) 099} [\href{https://arxiv.org/abs/1606.04549}{{\ttfamily
  1606.04549}}].

\bibitem{Cirigliano_2020}
V.~Cirigliano, W.~Detmold, A.~Nicholson and P.~Shanahan, \emph{Lattice {QCD}
  inputs for nuclear double beta decay},
  \href{https://doi.org/10.1016/j.ppnp.2020.103771}{\emph{Prog. Part. Nucl.
  Phys.} {\bfseries 112} (2020) 103771}
  [\href{https://arxiv.org/abs/2003.08493}{{\ttfamily 2003.08493}}].

\bibitem{chiral-eft-1}
S.~Weinberg, \emph{Nuclear forces from chiral {L}agrangians},
  \href{https://doi.org/10.1016/0370-2693(90)90938-3}{\emph{Phys. Lett. B}
  {\bfseries 251} (1990) 288}.

\bibitem{chiral-eft-2}
S.~Weinberg, \emph{Effective chiral {L}agrangians for nucleon-pion interactions
  and nuclear forces},
  \href{https://doi.org/10.1016/0550-3213(91)90231-L}{\emph{Nucl. Phys. B}
  {\bfseries 363} (1991) 3}.

\bibitem{chiral-eft-3}
S.~Weinberg, \emph{Three-body interactions among nucleons and pions},
  \href{https://doi.org/10.1016/0370-2693(92)90099-P}{\emph{Phys. Lett. B}
  {\bfseries 295} (1992) 114}.

\bibitem{Cirigliano_2018}
V.~Cirigliano, W.~Dekens, J.~de~Vries, M.L.~Graesser and E.~Mereghetti, \emph{A
  neutrinoless double beta decay master formula from effective field theory},
  \href{https://doi.org/10.1007/jhep12(2018)097}{\emph{JHEP} {\bfseries 12}
  (2018) 097} [\href{https://arxiv.org/abs/1806.02780}{{\ttfamily
  1806.02780}}].

\bibitem{nicholson}
A.~Nicholson et~al., \emph{Heavy physics contributions to neutrinoless double
  beta decay from {QCD}},
  \href{https://doi.org/10.1103/PhysRevLett.121.172501}{\emph{Phys. Rev. Lett.}
  {\bfseries 121} (2018) 172501}
  [\href{https://arxiv.org/abs/1805.02634}{{\ttfamily 1805.02634}}].

\bibitem{nplqcd-short-distance}
W.~Detmold, W.~Jay, D.~Murphy, P.~Oare and P.~Shanahan, \emph{Neutrinoless
  double beta decay from lattice {QCD}: The short-distance $\pi^- \rightarrow
  \pi^+ e^- e^-$ amplitude},
  \href{https://doi.org/10.1103/physrevd.107.094501}{\emph{Phys. Rev. D}
  {\bfseries 107} (2023) } [\href{https://arxiv.org/abs/2208.05322}{{\ttfamily
  2208.05322}}].

\bibitem{light-nuclei-nplqcd}
S.R.~Beane, E.~Chang, S.D.~Cohen, W.~Detmold, H.W.~Lin, T.C.~Luu et~al.,
  \emph{Light nuclei and hypernuclei from quantum chromodynamics in the limit
  of {SU}(3) flavor symmetry},
  \href{https://doi.org/10.1103/PhysRevD.87.034506}{\emph{Phys. Rev. D}
  {\bfseries 87} (2013) 034506}
  [\href{https://arxiv.org/abs/1206.5219}{{\ttfamily 1206.5219}}].

\bibitem{pp-fusion-lattice}
M.J.~Savage, P.E.~Shanahan, B.C.~Tiburzi, M.L.~Wagman, F.~Winter, S.R.~Beane
  et~al., \emph{Proton-proton fusion and tritium $\beta$-decay from lattice
  quantum chromodynamics},
  \href{https://doi.org/10.1103/PhysRevLett.119.062002}{\emph{Phys. Rev. Lett.}
  {\bfseries 119} (2017) 062002}
  [\href{https://arxiv.org/abs/1610.04545}{{\ttfamily 1610.04545}}].

\bibitem{nplqcd-2}
S.R.~Beane, E.~Chang, S.D.~Cohen, W.~Detmold, P.~Junnarkar, H.W.~Lin et~al.,
  \emph{Nucleon-nucleon scattering parameters in the limit of {SU}(3) flavor
  symmetry}, \href{https://doi.org/10.1103/physrevc.88.024003}{\emph{Phys. Rev.
  C} {\bfseries 88} (2013) } [\href{https://arxiv.org/abs/1301.5790}{{\ttfamily
  1301.5790}}].

\bibitem{nplqcd-4}
M.L.~Wagman, F.~Winter, E.~Chang, Z.~Davoudi, W.~Detmold, K.~Orginos et~al.,
  \emph{Baryon-baryon interactions and spin-flavor symmetry from lattice
  quantum chromodynamics},
  \href{https://doi.org/10.1103/physrevd.96.114510}{\emph{Phys. Rev. D}
  {\bfseries 96} (2017) } [\href{https://arxiv.org/abs/1706.06550}{{\ttfamily
  1706.06550}}].

\bibitem{nplqcd-variational}
S.~Amarasinghe, R.~Baghdadi, Z.~Davoudi, W.~Detmold, M.~Illa, A.~Parre{\~{n}}o
  et~al., \emph{Variational study of two-nucleon systems with lattice {QCD}},
  \href{https://doi.org/10.1103/physrevd.107.094508}{\emph{Phys. Rev. D}
  {\bfseries 107} (2023) } [\href{https://arxiv.org/abs/2108.10835}{{\ttfamily
  2108.10835}}].

\bibitem{nucleon-scattering}
E.~Berkowitz, T.~Kurth, A.~Nicholson, B.~Joo, E.~Rinaldi, M.~Strother et~al.,
  \emph{Two-nucleon higher partial-wave scattering from lattice {QCD}},
  \href{https://doi.org/10.1016/j.physletb.2016.12.024}{\emph{Phys. Lett. B}
  {\bfseries 765} (2017) 285}
  [\href{https://arxiv.org/abs/1508.00886}{{\ttfamily 1508.00886}}].

\bibitem{callat-variational}
B.~H{\"o}rz, D.~Howarth, E.~Rinaldi, A.~Hanlon, C.C.~Chang, C.~K{\"o}rber
  et~al., \emph{Two-nucleon $s$-wave interactions at the ${SU}(3)$
  flavor-symmetric point with $m_{ud} \simeq m_s^\text{phys}$: A first lattice
  {QCD} calculation with the stochastic {L}aplacian {H}eaviside method},
  \href{https://doi.org/10.1103/physrevc.103.014003}{\emph{Phys. Rev. C}
  {\bfseries 103} (2021) } [\href{https://arxiv.org/abs/2009.11825}{{\ttfamily
  2009.11825}}].

\bibitem{mainz-variational}
A.~Francis, J.R.~Green, P.M.~Junnarkar, C.~Miao, T.D.~Rae and H.~Wittig,
  \emph{Lattice {QCD} study of the {$H$} dibaryon using hexaquark and
  two-baryon interpolators},
  \href{https://doi.org/10.1103/physrevd.99.074505}{\emph{Phys. Rev. D}
  {\bfseries 99} (2019) } [\href{https://arxiv.org/abs/1805.03966}{{\ttfamily
  1805.03966}}].

\bibitem{background-field-1}
F.~Fucito, G.~Parisi and S.~Petrarca, \emph{First evaluation of {$g_A g_V$} in
  lattice {QCD} in the quenched approximation},
  \href{https://doi.org/10.1016/0370-2693(82)90816-4}{\emph{Phys. Lett. B}
  {\bfseries 115} (1982) 148}.

\bibitem{background-field-2}
G.~Martinelli, G.~Parisi, R.~Petronzio and F.~Rapuano, \emph{The proton and
  neutron magnetic moments in lattice {QCD}},
  \href{https://doi.org/10.1016/0370-2693(82)90162-9}{\emph{Phys. Lett. B}
  {\bfseries 116} (1982) 434}.

\bibitem{background-field-3}
C.~Bernard, T.~Draper, K.~Olynyk and M.~Rushton, \emph{Lattice
  quantum-chromodynamics calculation of some baryon magnetic moments},
  \href{https://doi.org/10.1103/PhysRevLett.49.1076}{\emph{Phys. Rev. Lett.}
  {\bfseries 49} (1982) 1076}.

\bibitem{2vbb-decay-lattice}
B.C.~Tiburzi, M.L.~Wagman, F.~Winter, E.~Chang, Z.~Davoudi, W.~Detmold et~al.,
  \emph{Double-$\beta$ decay matrix elements from lattice quantum
  chromodynamics},
  \href{https://doi.org/10.1103/PhysRevD.96.054505}{\emph{Phys. Rev. D}
  {\bfseries 96} (2017) 054505}
  [\href{https://arxiv.org/abs/1702.02929}{{\ttfamily 1702.02929}}].

\bibitem{sparsening}
W.~Detmold, D.J.~Murphy, A.V.~Pochinsky, M.J.~Savage, P.E.~Shanahan and
  M.L.~Wagman, \emph{Sparsening algorithm for multi-hadron lattice {QCD}
  correlation functions},
  \href{https://doi.org/10.1103/PhysRevD.104.034502}{\emph{Phys. Rev. D}
  {\bfseries 104} (2021) 034502}
  [\href{https://arxiv.org/abs/1908.07050}{{\ttfamily 1908.07050}}].

\bibitem{distillation-1}
J.~Foley, K.J.~Juge, A.{\'O}.~Cais, M.~Peardon, S.M.~Ryan and J.-I.~Skullerud,
  \emph{Practical all-to-all propagators for lattice {QCD}},
  \href{https://doi.org/10.1016/j.cpc.2005.06.008}{\emph{Comput. Phys. Comm.}
  {\bfseries 172} (2005) 145}
  [\href{https://arxiv.org/abs/hep-lat/0505023}{{\ttfamily hep-lat/0505023}}].

\bibitem{distillation-2}
{\scshape Hadron Spectrum Collaboration} collaboration, \emph{Novel quark-field
  creation operator construction for hadronic physics in lattice {QCD}},
  \href{https://doi.org/10.1103/PhysRevD.80.054506}{\emph{Phys. Rev. D}
  {\bfseries 80} (2009) 054506}
  [\href{https://arxiv.org/abs/0905.2160}{{\ttfamily 0905.2160}}].

\bibitem{distillation-3}
C.~Morningstar, J.~Bulava, J.~Foley, K.J.~Juge, D.~Lenkner, M.~Peardon et~al.,
  \emph{Improved stochastic estimation of quark propagation with {L}aplacian
  {H}eaviside smearing in lattice {QCD}},
  \href{https://doi.org/10.1103/PhysRevD.83.114505}{\emph{Phys. Rev. D}
  {\bfseries 83} (2011) 114505}
  [\href{https://arxiv.org/abs/1104.3870}{{\ttfamily 1104.3870}}].

\bibitem{0vbb-paper}
Z.~Davoudi, W.~Detmold, Z.~Fu, A.V.~Grebe, W.~Jay, D.~Murphy et~al.,
  \emph{Long-distance nuclear matrix elements for neutrinoless double-beta
  decay from lattice {QCD}},
  \href{https://arxiv.org/abs/2402.09362}{{\ttfamily 2402.09362}}.

\bibitem{zohreh-nu-prop}
Z.~Davoudi and S.V.~Kadam, \emph{The path from lattice {QCD} to the
  short-distance contribution to $0\nu\beta\beta$ decay with a light {M}ajorana
  neutrino}, \href{https://doi.org/10.1103/physrevlett.126.152003}{\emph{Phys.
  Rev. Lett.} {\bfseries 126} (2021) }
  [\href{https://arxiv.org/abs/2012.02083}{{\ttfamily 2012.02083}}].

\bibitem{zi-yu-presentation}
Z.-Y.~Wang, \emph{Lattice calculation of proton-proton fusion matrix element},
  {\emph{Proc. Sci.} {\bfseries LATTICE2024} (2025) 111}.

\bibitem{davoudi-2}
Z.~Davoudi and S.V.~Kadam, \emph{On the extraction of low-energy constants of
  single- and double-$\beta$ decays from lattice {QCD}: A sensitivity
  analysis}, \href{https://doi.org/10.1103/PhysRevD.105.094502}{\emph{Phys.
  Rev. D} {\bfseries 105} (2022) 094502}
  [\href{https://arxiv.org/abs/2111.11599}{{\ttfamily 2111.11599}}].

\bibitem{multigrid-1}
J.~Brannick, R.C.~Brower, M.A.~Clark, J.C.~Osborn and C.~Rebbi, \emph{Adaptive
  multigrid algorithm for lattice {QCD}},
  \href{https://doi.org/10.1103/PhysRevLett.100.041601}{\emph{Phys. Rev. Lett.}
  {\bfseries 100} (2008) 041601}
  [\href{https://arxiv.org/abs/0707.4018}{{\ttfamily 0707.4018}}].

\bibitem{multigrid-2}
R.~Babich, J.~Brannick, R.C.~Brower, M.A.~Clark, T.A.~Manteuffel,
  S.F.~McCormick et~al., \emph{Adaptive multigrid algorithm for the lattice
  {W}ilson-{D}irac operator},
  \href{https://doi.org/10.1103/PhysRevLett.105.201602}{\emph{Phys. Rev. Lett.}
  {\bfseries 105} (2010) 201602}
  [\href{https://arxiv.org/abs/1005.3043}{{\ttfamily 1005.3043}}].

\bibitem{multigrid-3}
J.C.~Osborn, R.~Babich, J.~Brannick, R.C.~Brower, M.A.~Clark, S.D.~Cohen
  et~al., \emph{Multigrid solver for clover fermions},
  \href{https://doi.org/10.22323/1.105.0037}{\emph{PoS} {\bfseries LATTICE2010}
  (2010) 037} [\href{https://arxiv.org/abs/1011.2775}{{\ttfamily 1011.2775}}].

\bibitem{parisi}
G.~Parisi, \emph{The strategy for computing the hadronic mass spectrum},
  \href{https://doi.org/10.1016/0370-1573(84)90081-4}{\emph{Phys. Rep.}
  {\bfseries 103} (1984) 203}.

\bibitem{lepage}
G.P.~Lepage, \emph{The analysis of algorithms for lattice field theory},  in
  \emph{From Actions to Answers. Proceedings, {T}heoretical {A}dvanced {S}tudy
  {I}nstitute in Elementary-Particle Physics}, T.A.~DeGrand and D.~Toussaint,
  eds., (Singapore), pp.~97--121, World Scientific (1990).

\end{thebibliography}\endgroup

\end{document}